\RequirePackage{snapshot}
\documentclass[10pt,conference]{IEEEtran}

\usepackage{graphicx}
\usepackage[cmex10]{amsmath}
\usepackage{amsfonts}
\usepackage[caption=false,font=footnotesize]{subfig}

\newtheorem{theorem}{Theorem}
\newtheorem{lemma}[theorem]{Lemma}
\newtheorem{proposition}[theorem]{Proposition}

\newtheorem{remark}{Remark}

\newcommand\goesto{\rightarrow}
\newcommand\ra{\rightarrow}
\newcommand{\toinf}[1]{\ensuremath{#1 \goesto\infty}}
\newcommand{\limtoinf}[1]{\ensuremath{\lim_{\toinf{#1}}}}
\newcommand{\Sh}{{\hat{S}}}
\newcommand{\Qh}{{\hat{Q}}}
\newcommand{\Eh}{{\hat{E}}}

\newcommand{\N}{{\mathcal{N}}}
\newcommand{\E}{\mathbb{E}}
\newcommand{\R}{\mathbb{R}}
\newcommand{\Z}{\mathbb{Z}}
\newcommand{\I}{{\mathcal{I}}}

\newcommand{\ssq}{{\sigma_S^2}}
\newcommand{\szq}{{\sigma_Z^2}}
\newcommand{\seq}{{\sigma_E^2}}

\newcommand\eg{\emph{e.g.}}

\newcommand\e{\epsilon}

\newcommand\snr{\textsc{snr}}

\newcommand\sdr{\textsc{sdr}}

\newcommand{\Err}{{\mathcal{E}}}
\newcommand{\Eq}{{\Err_Q}}
\newcommand{\Eqi}{{\Err_{Q,i}}}
\newcommand{\Ee}{{\Err_E}}

\newcommand\deq{\stackrel{\mathrm{def}}{=}}

\DeclareMathOperator\Int{int}
\DeclareMathOperator\var{Var}

\begin{document}

\title{Asymptotically Optimal Joint Source-Channel Coding with Minimal Delay}
\author{\IEEEauthorblockN{Marius Kleiner, Bixio Rimoldi}
\IEEEauthorblockA{School of Computer and Communication Sciences\\
Ecole Polytechnique F\'ed\'erale de Lausanne\\
CH-1015 Lausanne, Switzerland\\
E-mail: firstname.lastname@epfl.ch}}

\maketitle

\begin{abstract}
We present and analyze a joint source-channel coding strategy for the
transmission of a Gaussian source across a Gaussian channel in $n$~channel
uses per source symbol. Among all such strategies, the scheme presented here has
the following properties: i)~the resulting mean-squared error scales optimally
with the signal-to-noise ratio, and ii)~the scheme is easy to implement and the
incurred delay is minimal, in the sense that a single source symbol is encoded
at a time. 
\end{abstract}

\section{Introduction}

In this paper we propose and analyze a scheme for the transmission of a
discrete-time memoryless Gaussian source across a discrete-time memoryless
Gaussian channel, where the channel can be used $n$ times for each source
symbol. The parameter~$n$ is arbitrary but fixed, given as part of the problem
statement.

It is well known that if the source has variance~$\ssq$ and the channel noise
has variance~$\szq$ then the average transmit power~$P$ and the average
mean-squared error $D$ of any communication scheme for this scenario are related
by
\begin{equation}
  \label{eq:shannonlimit}
  R(D) \le nC(P),
\end{equation}
where $R(D) = 0.5 \log(\ssq/D)$ is the rate-distortion function of the source
and $C(P) = 0.5 \log( 1 + P/\szq)$ is the capacity-cost function of the channel
(see \eg~\cite{CoverT1991}).
Inserting into~\eqref{eq:shannonlimit} yields
\begin{equation*}
  \frac{\ssq}{D} \le \left( 1 + \frac{P}{\szq} \right)^n, 
\end{equation*}
or equivalently
\begin{equation}
  \label{eq:sdrbound}
  \sdr \le (1 + \snr)^n,
\end{equation}
where we have defined $\snr = P/\szq$ and $\sdr = \ssq/D$.
In the limit when \snr\ goes to infinity, 
\begin{equation}
  \label{eq:scalingbound}
  \limtoinf{\snr} \frac{\log\sdr}{\log\snr} \le   n.
\end{equation}
At large SNR, the SDR (signal-to-distortion ratio) behaves thus at best as
$\snr^n$. In this sense $n$ is the best possible scaling exponent that any
communication scheme can hope to achieve for a fixed~$n$.

The scheme proposed in this paper achieves this optimal scaling exponent for any
fixed~$n$, yet has small complexity and minimal delay in the sense that it
operates on a single source symbol at a time. It works by quantizing the source
and then repeatedly quantizing the quantization error. The quantized points are
sent across the first $n-1$ channel uses and the last quantization error is
sent uncoded in the $n^{\mathrm{th}}$ channel use.


If the quantization resolution is chosen correctly (as a function of the SNR),
then the decoding error of the quantization symbols is dominated by that of the
uncoded transmission in the last channel use, which is shown to have an
asymptotic scaling exponent of~$n$.

\begin{figure}
  \centerline{\subfloat[Shannon's original proposition.]{\includegraphics{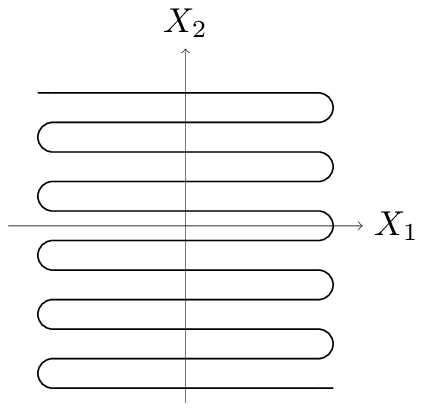}}
  \hfil
  \subfloat[Mapping proposed in this paper
  (for~$n=2$).]{\includegraphics{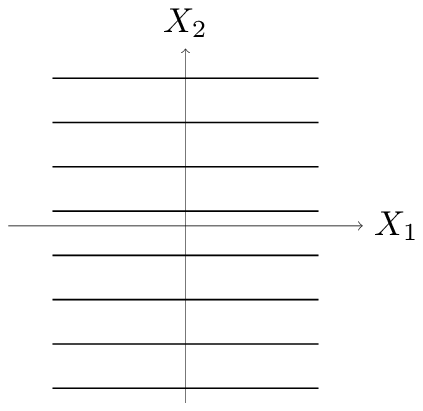}} }
  \caption{A minimum-delay source-channel code for $n=2$ can be visualized as a
  curve in $\R^2$ parametrized by the source. Here we compare the mapping
  presented in this paper (right) to Shannon's original suggestion (left).}
  \label{fig:shannoncomparison}
\end{figure}

Schemes similar to the one proposed here have been considered before. Indeed,
one of the first schemes to transmit an analog source across two uses of a
Gaussian channel was suggested by Shannon~\cite{Shannon1949}. Notice its
resemblance to the constellation studied here, shown in
Figure~\ref{fig:shannoncomparison}.

After Shannon, Wozencraft and Jacobs~\cite{WozencraftJ1965} were among the first
to study source-channel mappings as curves in $n$-dimensional space.
Ziv~\cite{Ziv1970} found important theoretical limitations of such mappings.
Much of the later work is due to Ramstad and his coauthors
(see~\cite{Ramstad2002}, \cite{FloorR2006}, \cite{CowardR2000,CowardR2000a},
\cite{WernerssonSR2007}, \cite{HeklandFR2009}). A proof that the performance of
minimal-delay codes is strictly smaller than that of codes with unrestricted
delay when $n>1$ was given in 2008 by Ingber et al.~\cite{IngberLZF2008}.

For $n=2$, the presented scheme is almost identical to the HSQLC scheme by
Coward~\cite{Coward2001}, which uses a numerically optimized quantizer,
transmitter and receiver to minimize the mean-squared error (MSE) for finite
values of the SNR. Coward correctly conjectured that the right strategy for $n >
2$ would be to repeatedly quantize the quantization error from the previous
step, which is exactly what we do here.

Another closely related communication scheme is the \emph{shift-map} scheme due
to Chen and Wornell~\cite{ChenW1998}.  Vaishampayan and
Costa~\cite{VaishampayanC2003} showed in their analysis that it achieves the
scaling exponent $n-\e$ for any $\e > 0$ if the relevant parameters are chosen
correctly as a function of the SNR. Up to rotation and a different constellation
shaping, the shift-map scheme is in fact virtually identical to the one
presented here, a fact that was pointed out recently by Taherzadeh and
Khandani~\cite{TaherzadehK2008}. In their own paper they develop a scheme that
achieves the optimal scaling exponent exactly and is in addition robust to SNR
estimation errors; their scheme, however, is based on rearranging the digits of
the binary expansion of the source and is thus quite different from the one
presented here.

Shamai, Verd\'u and Zamir~\cite{ShamaiVZ1998} used Wyner-Ziv coding to extend an
existing analog system with a digital code when additional bandwidth is
available. Mittal and Phamdo~\cite{MittalP2002} (see also the paper by Skoglund,
Phamdo and Alajaji~\cite{SkoglundPA2002}) split up the source into a quantized
part and a quantization error, much like we do here, but they use a
separation-based code (or ``tandem'' code) to transmit the quantization symbols.
Reznic et al.~\cite{ReznicFZ2006} use both quantization and Wyner-Ziv coding,
and their scheme includes Shamai et al.\ and Mittal \& Phamdo as extreme cases.
All three schemes, however, use long block codes for the digital phase and incur
correspondingly large delays, so they are not directly comparable with minimum
delay schemes.

While the basic idea of the scheme considered in this paper is not new, the
analysis provided is and, to our knowledge, we are the first to give an
exact mathematical formulation of the quantization resolution (as a function of
the SNR) that leads to the optimal scaling exponent.

%

\section{Proposed Communication Scheme}

\subsection{Encoder}
To encode a single source letter $S$ into $n$~channel input symbols $X_1$,
\dots, $X_n$, we proceed as follows. Define $E_0 = S$ and recursively compute
the pairs $(Q_i, E_i)$ as
\begin{align}
  Q_i &= \frac{1}{\beta} \Int(\beta E_{i-1}) \nonumber \\
  E_i &= \beta (E_{i-1} - Q_i) \label{eq:QEdef}
\end{align}
for $i = 1$, \dots, $n-1$ where $\Int(x)$ is the unique integer~$i$ such that
\begin{equation*}
  x \in \left[ i - \frac12 , i + \frac12 \right)
\end{equation*}
and $\beta$ is a scaling factor that \emph{grows with the power $P$}
in a way to be determined later.  The following result will be useful in the
sequel.
\begin{lemma}
  \label{lem:qvarconvergence}
  As $\beta$ goes to infinity, the variance of $Q_i$ converges to that of
  $E_{i-1}$ for all $i = 1$, \dots, $n-1$. 
\end{lemma}
\medskip

\begin{IEEEproof}
  Intuitively this is so since $Q_i$ is $E_{i-1}$ quantized, and the
  quantization step becomes smaller as $\beta$ goes to infinity. A rigorous
  proof is given in Appendix~\ref{app:lemma1proof}.
\end{IEEEproof}

\begin{proposition}
  \label{prop:qeproperties}
  The $Q_i$ and $E_i$ satisfy the following properties:
\begin{enumerate}
  \item The map $S \mapsto (Q_1, \dots, Q_{n-1}, E_{n-1})$ is one-to-one and
    \begin{equation}
      \label{eq:unwraprec}
      S = \sum_{i=1}^{n-1} \frac{1}{\beta^{i-1}} Q_i + \frac{1}{\beta^{n-1}}
      E_{n-1}.
    \end{equation}

  \item The variance of $E_0$ is $\ssq$ and for all $i = 1$, \dots, $n-1$, $E_i
    \in [-1/2, 1/2)$ and $\var(E_i) \le 1/4$.
  \item For any $\delta > 0$ there exists $\beta_0$ such that for $\beta
    > \beta_0$,
    \begin{equation}
      \label{eq:qvarbound}
      \var(Q_i) \le
      \begin{cases}
        \ssq + \delta & \text{for $i = 1$} \\
        1/4 + \delta & \text{for $i = 2$, \dots, $n-1$.}
      \end{cases}
    \end{equation}
\end{enumerate}
\end{proposition}

\goodbreak
\begin{IEEEproof}
  \begin{enumerate}
    \item From the definition~\eqref{eq:QEdef} we have
    \begin{equation}
      \label{eq:reverserec}
      E_{i-1} = \frac{1}{\beta} E_i + Q_i.
    \end{equation}
    Repeated use of this relationship leads to the given expression for~$S$. 
  \item First, $\var(E_0) = \var(S) = \ssq$. Next, $E_i \in [-1/2, 1/2)$ follows
    trivially from the definition of~$E_i$.  Furthermore, the variance of any
    random variable with support in an interval of length~$1$ is bounded from
    above by~$1/4$. 
  \item The result follows directly from Lemma~\ref{lem:qvarconvergence} and
    from the bound on the variance of the $E_i$ in point~2 above.
  \end{enumerate}
\end{IEEEproof}

Without loss of generality we assume hereafter that $\ssq > 1/4$ so that the
first bound of~\eqref{eq:qvarbound} applies to all~$Q_i$.

We determine the channel input symbols $X_i$ from the $Q_i$ and from $E_{n-1}$
according to 
\begin{align*}
  X_i &= \sqrt{\frac{P}{\ssq + \delta}} Q_i \quad \text{for $i = 1$,
  \dots, $n-1$
  and} \\
  X_n &= \sqrt{\frac{P}{\seq}} E_{n-1},
\end{align*}
where $\seq = \var(E_{n-1})$.  Following Proposition~\ref{prop:qeproperties},
this ensures that $\E[X_i^2] \le P$ for all~$i$ and for $\beta >
\beta_0(\delta)$.  Since we are interested in the large SNR regime and since we
have defined $\beta$ to grow with~$P$, we can thus assume for the remainder that
the power constraint is satisfied.

\subsection{Decoder}

The $X_i$ are transmitted across the channel, producing at the channel output
the symbols
\begin{equation*}
  Y_i = X_i + Z_i, \quad i = 1, \dots, n,
\end{equation*}
where the $Z_i$ are iid Gaussian random variables of variance~$\szq$. 
To estimate $S$ from  $Y_1$, \dots, $Y_n$, the decoder first
computes separate estimates $\Qh_1$, \dots, $\Qh_{n-1}$ and $\Eh_{n-1}$, and
then combines them to obtain the final estimate~$\Sh$.  While this strategy is
suboptimal in terms of achieving a small MSE, we will see that it is good enough
to achieve optimal scaling.

To estimate the $Q_i$ we use a maximum likelihood (ML) decoder, which yields the
minimum distance estimate
\begin{equation}
  \label{eq:mldecoder}
  \Qh_i = \frac{1}{\beta} \arg \min_{j\in \Z} \left| \sqrt{\frac{P}{\ssq
  + \delta} } \frac{j}{\beta} - Y_i \right|.
\end{equation}
To estimate $E_{n-1}$, we use a linear minimum mean-square error (LMMSE)
estimator (see \eg~\cite[Section~8.3]{Scharf1990}), which computes
\begin{equation}
  \label{eq:lmmse}
  \Eh_{n-1} = \frac{\E[E_{n-1} Y_n]}{\E[Y_n^2]} Y_n.
\end{equation}
Finally we use the relationship~\eqref{eq:unwraprec} to obtain
\begin{equation}
  \label{eq:unwrapestim}
  \Sh = \sum_{i=1}^{n-1} \frac{1}{\beta^{i-1}} \Qh_i + \frac{1}{\beta^{n-1}}
  \Eh_{n-1}.
\end{equation}

\subsection{Error Analysis}

The overall MSE $\E[(S-\Sh)^2]$ can be broken up into contributions due to the
errors in decoding $Q_i$ and $E_{n-1}$ as follows. From~\eqref{eq:unwraprec}
and~\eqref{eq:unwrapestim}, the difference between $S$ and $\Sh$ is
\begin{equation*}
  S - \Sh = \sum_{i=1}^{n-1} \frac1{\beta^{i-1}} (Q_i - \Qh_i) + \frac1{\beta^{n-1}}
  (E_{n-1} - \Eh_{n-1}).
\end{equation*}
The error terms $Q_i - \Qh_i$ depend only on the noise of the respective channel
uses and are therefore independent of each other and of $E_{n-1} - \Eh_{n-1}$,
so we can write the error variance componentwise as
\begin{equation}
  \label{eq:totalerror}
  \E[(S-\Sh)^2] = \sum_{i=1}^{n-1} \frac{1}{\beta^{2(i-1)}} \Eqi +
  \frac{1}{\beta^{2(n-1)}} \Ee, 
\end{equation}
where $\Eqi \deq \E[(Q_i - \Qh_i)^2]$ and $\Ee \deq \E[(E_{n-1} -
\Eh_{n-1})^2]$.

\begin{lemma}
  \label{lem:eqbound}
  For each $i = 1$, \dots, $n-1$, 
  \begin{equation}
    \label{eq:eqidecay}
    \Eqi \in O\left(\exp\{-k \snr/\beta^2\}\right),
  \end{equation}
  where $\snr = P/\szq$ and $k > 0$~is a constant.
\end{lemma}
(The $O$-notation is defined in Appendix~\ref{app:Onotation}.)

\begin{IEEEproof}
  Define the interval
  \begin{equation*}
    \I_j = \left[ \frac{(j - \frac12) \sqrt{P}}{\beta \sqrt{\ssq + \delta}},
    \frac{(j + \frac12) \sqrt{P}}{\beta \sqrt{\ssq + \delta}} \right).
  \end{equation*}
  According to the minimum distance decoder~\eqref{eq:mldecoder}, $\Qh_i - Q_i
  = j/\beta$ whenever $Z_i \in \I_j$.  The error $\Eqi$ satisfies thus
  \begin{align}
    \E[(Q_i - \Qh_i)^2] &= \frac{1}{\beta^2} \sum_{j \in \Z} j^2 \Pr[Z_i \in
    \I_j]  \nonumber \\
    &= \frac{2}{\beta^2} \sum_{j = 1}^\infty j^2 \Pr[Z_i \in \I_j], \label{eq:eqexact}
  \end{align}
  where the second equality follows from the symmetry of the distribution
  of~$Z_i$. Now,
  \begin{equation*}
    \Pr[Z_i \in \I_j] = Q\left( \frac{(j - \frac12) \sqrt \snr}{\beta \sqrt{\ssq
    + \delta }} \right) - Q\left( \frac{(j + \frac12) \sqrt{\snr}}{\beta
    \sqrt{\ssq + \delta}} \right),
  \end{equation*}
  where
  \begin{equation*}
    Q(x) = \frac{1}{\sqrt{2\pi}} \int_x^\infty e^{-\xi^2/2} d\xi,
  \end{equation*}
  which can be bounded from above for $x \ge 0$ as
  \begin{equation*}
    Q(x) \le \frac12 e^{-x^2/2}.
  \end{equation*}
  For $\beta \ge 1$ we can now bound~\eqref{eq:eqexact} as
  \begin{equation*}
    \Eqi \le \sum_{j=1}^\infty j^2 \exp\left\{ - \frac{(j - 1/2)^2
    \snr}{2\beta^2(\ssq + \delta)} \right\}.
  \end{equation*}
  Note that for $j \ge 2$, $(j - 1/2)^2 > j$.  Thus
  \begin{eqnarray}
    \Eqi &\le & \exp \left \{ - \frac{\snr}{8 \beta^2 (\ssq +\delta)} \right\}
    \nonumber \\
    & & \mbox{} + 
    \sum_{j = 2}^\infty j^2 \exp \left \{ - \frac{j \snr}{2\beta^2(\ssq
    +\delta)}
    \right\}. \label{eq:eqibound}
  \end{eqnarray}
  To bound the infinite sum we use 
  \begin{equation}
    \label{eq:geomsum}
    \sum_{j=2}^\infty j^2 p^j \le \sum_{j=1}^\infty j^2 p^j = 
    \frac{p^2+p}{(1-p)^3}
  \end{equation}
  with $p = \exp\{-\snr/2 \beta^2 (\ssq+\delta)\}$. The first term
  of~\eqref{eq:eqibound} thus dominates for large values of
  $\snr/\beta^2$ and
  \begin{equation*}
    \Eqi \le c\exp\left\{ - \frac{\snr}{8 \beta^2 (\ssq + \delta)} \right\}
  \end{equation*}
  for some~$c > 0$, which completes the proof. 
\end{IEEEproof}

\begin{lemma}
  \label{lem:eedecay}
  $\Ee \in O(\snr^{-1})$. 
\end{lemma}
\begin{IEEEproof}
  The mean-squared error that results from the LMMSE estimation~\eqref{eq:lmmse}
  is
  \begin{equation}
    \label{eq:lmmse-error}
    \Ee = \seq - \frac{(\E[E_{n-1}
    Y_n])^2}{\E[Y_n^2]}. 
  \end{equation}
  Since
  \begin{equation*}
    Y_n = X_n + Z_n = \sqrt{\frac{P}{\seq}} E_{n-1} + Z_n,
  \end{equation*}
  we have $\E[E_{n-1}Y_n] = \sqrt{P\seq}$. Moreover, $\E[Y_n^2] = \E[X^2] +
  \E[Z^2] = P+\szq$.  Inserting this into~\eqref{eq:lmmse-error} we obtain
  \begin{align*}
    \Ee &= \seq - \frac{P \seq}{P + \szq} \\
    &= \seq \left( 1 - \frac{P}{P + \szq} \right) \\
    &= \frac{\seq}{1 + \snr} \\
    & < \frac{\seq}{\snr}.
  \end{align*}
  Since $\seq$ is bounded (cf.\ Proposition~\ref{prop:qeproperties}), 
  $\Ee \in O(\snr^{-1})$ as claimed.
\end{IEEEproof}

\subsection{Achieving the Optimal Scaling Exponent}

Recall the formula for the overall error
\begin{equation*}
  \E[(S-\Sh)^2] = \sum_{i=1}^{n-1} \frac{1}{\beta^{2(i-1)}} \Eqi +
  \frac{1}{\beta^{2(n-1)}} \Ee.
\end{equation*}
According to Lemma~\ref{lem:eqbound}, $\Eqi$ decreases exponentially
when $\snr/\beta^2$ goes to infinity. This happens for increasing \snr\ if we
set \eg
\begin{equation*}
  \beta^2 = \snr^{1-\e}
\end{equation*}
for some $\e > 0$, in which case $\Eqi \in O\left(\exp(-k \snr^\e) \right)$.
From this and Lemma~\ref{lem:eedecay}, the overall error satisfies
\begin{equation}
  \label{eq:overallO}
  \E[(S-\Sh)^2] \in O(\snr^{-(n - \e')}),
\end{equation}
where $\e' = (n-1)\e$ can be made as small as desired. The scaling exponent for
a fixed $\e$ satisfies therefore
\begin{equation}
  \label{eq:sdrepsilon}
  \limtoinf{\snr} \frac{\log\sdr}{\log\snr} \ge
  \limtoinf{\snr} \frac{\log \ssq + (n - \e') \log\snr}{\log \snr} = n - \e'. 
\end{equation}

Note that the choice of $\e$ represents a tradeoff: for small $\e$ the error due
to the ``discrete'' part vanishes only slowly, but the scaling exponent in the
limit is larger. For larger $\e$, $\Eq$ vanishes quickly but the resulting
exponent is smaller. In the remainder of this section we show how we can choose
$\e$ as a function of~$\snr$ to achieve the optimal scaling. 

Let now
\begin{equation}
  \label{eq:esnrdecay}
  \e = \e(\snr) = \frac{\log(n \log\snr / k)}{\log\snr},
\end{equation}
where $k$ is the constant indicating the decay of $\Eqi$ in~\eqref{eq:eqidecay}.
With this choice of $\e$,
\begin{align*}
  \Eqi &\in O\left( \exp\left( - k \snr^\e \right) \right) \\
 &= O(\snr^{-n}),
\end{align*}
hence the overall error is still dominated as in~\eqref{eq:overallO},
and~\eqref{eq:sdrepsilon} still applies. Inserting~\eqref{eq:esnrdecay}
in~\eqref{eq:sdrepsilon}, we find
\begin{align*}
  \lefteqn{ \limtoinf{\snr} \frac{\log\ssq + (n-(n-1)\e) \log\snr}{\log\snr} }
  \quad \\
  &= \limtoinf{\snr} \frac{\log\ssq + n\log\snr - (n-1)\log(n\log\snr/k)}
  {\log\snr} \\
  &= n,
\end{align*}
which is indeed the optimal scaling exponent.

\begin{remark}
  While the limiting \emph{exponent} above is indeed the optimal one, the SDR
  scales as $\snr^n (\log\snr)^{-(n-1)}$  rather than the theoretic optimum
  $\snr^n$.  This means that the \emph{gap} (in dB) between the theoretically
  optimal $\sdr$ value and our lower bound grows to infinity as $\snr \ra
  \infty$.  According to a result in~\cite{TaherzadehK2008}, however, no scheme
  combining quantization and uncoded transmission as done here can achieve a
  better SDR scaling than $\snr^n (\log\snr)^{-(n-1)}$.  In the scaling sense,
  our bound is therefore tight.
\end{remark}

\section{Conclusions}

We have presented and analyzed a joint source-channel communication strategy
that achieves the optimal scaling exponent if the channel is to be used
$n$~times per source symbol. The given scheme incurs the smallest possible
delay and its implementation is straightforward.

While the basic structure of this scheme -- separating the source into a
quantized part and the associated error -- is not new, the simple analysis
provided here yields an explicit expression for the quantization
resolution in terms of the SNR that leads to the optimal scaling exponent.


\appendices

\section{Proof of Lemma~\ref{lem:qvarconvergence}}
\label{app:lemma1proof}

Since all involved distributions are symmetric, $\E[Q_i] = 0$.  Writing $Q_i$ as
a function of $E_{i-1}$, we have
\begin{equation}
  \var(Q_i) = \E[Q_i^2] = \int_{-\infty}^\infty Q_i(\xi)^2 f(\xi) d\xi,
  \label{eq:varqintegral}
\end{equation}
where $f(\xi)$ is the pdf\footnote{probability density function} of $E_{i-1}$.
Now, $Q_i(\xi) = j/\beta$ whenever
\begin{equation*}
  \xi \in \left[ \frac{j - 1/2}{\beta}, \frac{j + 1/2}{\beta} \right).
\end{equation*}
With this, the integral~\eqref{eq:varqintegral} becomes
\begin{align*}
  \var(Q_i) &= \frac{1}{\beta^2} \sum_{j \in \Z} j^2 
  \int_{\frac{j - 1/2}{\beta}}^{\frac{j + 1/2}{\beta}} f(\xi) d\xi \\
  &= \sum_{j\in \Z} \left( \frac{j}{\beta} \right)^2 \left[ F\left( 
  { \textstyle
  \frac{j + 1/2}{\beta} }\right) - F \left( { \textstyle \frac{j - 1/2}{\beta}
  } \right) \right],
\end{align*}
where $F(\xi)$ is the cdf\footnote{cumulative distribution function} of $E_{i-1}$. As $\beta$
goes to infinity, this sum converges to a Riemann-Stieltjes integral:
\begin{equation*}
  \var(Q_i) \longrightarrow \int \xi^2 dF(\xi) = \var(E_{i-1}) \quad
  \text{as $\beta \goesto \infty$.}
\end{equation*}
\hfill\IEEEQED

\section{Big-O Notation}
\label{app:Onotation}

The ``Big-O'' asymptotic notation used at various points in the paper is defined
as follows.  Let $f(x)$ and $g(x)$ be two functions defined on~$\R$. We write
\begin{equation*}
  f(x) \in O(g(x))
\end{equation*}
if and only if there exists an $x_0$ and a constant~$c$ such that
\begin{equation*}
  f(x) \le c g(x)
\end{equation*}
for all $x > x_0$. 

As a simple consequence of this definition, if $f(x) \in O(x^n)$ then 
\begin{equation*}
  \limtoinf{x} \frac{\log f(x)}{\log x} \le n.
\end{equation*}

\bibliography{mkbiblio}
\end{document}